\documentclass{article}

\usepackage[margin=1in]{geometry}

\usepackage[
style=authoryear-icomp,
maxbibnames=9,
maxcitenames=2,
backend=biber]{biblatex}

\usepackage{graphicx}
\usepackage[colorlinks,
citecolor=blue,
linkcolor=blue,
urlcolor=blue
]{hyperref}
\usepackage{array}
\usepackage{booktabs}
\usepackage{comment}
\usepackage{subcaption}

\usepackage{todonotes}

\usepackage{caption}
\usepackage[page]{appendix} 

\usepackage{authblk}

\title{Comparing Traditional and LLM-based Search for Consumer Choice: A Randomized Experiment}

\author{Sofia Eleni Spatharioti}
\author{David M. Rothschild}
\author{Daniel G. Goldstein}
\author{Jake M. Hofman}

\affil{Microsoft Research, New York City}
\date{}                    
\begin{document}

\maketitle

\begin{abstract}

Recent advances in the development of large language models are rapidly
changing how online applications function. LLM-based search tools, for
instance, offer a natural language interface that can accommodate
complex queries and provide detailed, direct responses. At the same
time, there have been concerns about the veracity of the information
provided by LLM-based tools due to potential mistakes or fabrications
that can arise in algorithmically generated text. In a set of online
experiments we investigate how LLM-based search changes people's
behavior relative to traditional search, and what can be done to
mitigate overreliance on LLM-based output. Participants in our
experiments were asked to solve a series of decision tasks that involved
researching and comparing different products, and were randomly assigned
to do so with either an LLM-based search tool or a traditional search
engine. In our first experiment, we find that participants using the
LLM-based tool were able to complete their tasks more quickly, using
fewer but more complex queries than those who used traditional search.
Moreover, these participants reported a more satisfying experience with
the LLM-based search tool. When the information presented by the LLM was
reliable, participants using the tool made decisions with a comparable
level of accuracy to those using traditional search; however we observed
overreliance on incorrect information when the LLM erred. Our second
experiment further investigated this issue by randomly assigning some
users to see a simple color-coded highlighting scheme to alert them to
potentially incorrect or misleading information in the LLM responses.
Overall we find that this confidence-based highlighting substantially
increases the rate at which users spot incorrect information, improving
the accuracy of their overall decisions while leaving most other
measures unaffected. Together these results suggest that LLM-based
information retrieval tools have promise for increasing the productivity
of people engaged in decision tasks, and highlights the opportunity of
communicating uncertainty to help people know when to do further
research.

\end{abstract}

\section{Introduction}\label{sec:introduction}

Recent advances in artificial intelligence (AI), specifically in large language models (LLMs), are changing the tools that billions of people use in their daily lives.
One of the first applications to be transformed was the search engine.
ChatGPT, an LLM chatbot, was released on November 30, 2022, and by February 2023, Microsoft and Google announced the upcoming availability of LLM-based search engines and began a rapid rollout with Microsoft ending its waitlist for Bing Chat on May 4, 2023.

From a user experience perspective, traditional web search and LLM-based search differ in a number of ways, each having their own advantages and disadvantages.
When using traditional web search, users typically issue relatively succinct queries~(\cite{silverstein1999analysis,jansen2000real}) and are presented with a list of hyperlinks to and snippets from web pages containing relevant reference information.
There are several benefits of this style of information retrieval. 
First, traditional search allows rather direct access to source material through hyperlinks.
Second, traditional search enables users to see convergence or disagreement among distinct sources of information through the different references on a results page.
Third, traditional search is explicitly optimized to return authoritative results~(\cite{brin1998pagerank}) and provides additional cues about the reliability of information, for example through the domains and publishers of different results (e.g., information from the Library of Congress might be considered more trustworthy than one from an unknown domain).

At the same time, there are some drawbacks to the traditional web search process.
While it is convenient to have access to reference material from different sources, synthesizing information from them can be challenging and time consuming.
Whereas relevant information is sometimes presented in the snippets or ``instant answers'' on a search result page, users often have to click through to several different results and search within those respective pages to find pertinent information.
In addition, verbose or complex queries can often lead to poor search results~(\cite{bendersky2008discovering,gupta2015information}), and given that many real-world decision tasks are complex, this can result in users needing to break down a task into a series of simpler queries~(\cite{jiang2014searching}). 
Lastly, it can be a technical challenge for search engines to retain context among sequences of such queries within a complex search session~(\cite{lawrence2000context,finkelstein2001placing}).

LLM-based search has a different set of strengths and weaknesses.
In terms of strengths, LLM-based search offers a natural language interface than can handle complex queries and return detailed, direct responses.
This includes the ability to extract details from within many different references and synthesize potentially complex information across them.
LLM-based search also lends itself to retaining more context that traditional search, allowing users to engage in a conversational exchange to refine and follow up on a sequence of queries.

At the same time, LLM-based search currently faces a number of challenges. LLMs are known to have issues with ``fabrication'' or ``hallucination'' in which they generate plausible sounding but factually inaccurate sequences of text~(\cite{maynez2020faithfulness}).
This is particularly worrisome in the context of using LLMs for web search, with the concern that fabrications might lead to overreliance on incorrect search results if users simply assume that what they are shown is always correct.
Furthermore, compared with traditional search, LLM-based search offers fewer reliable cues for users to gauge the accuracy of information.
Responses may lack hyperlinks to source materials, which users rely on to verify statements.
Even when external links are provided, they are not as prominent as in traditional web search, often appearing as subtle footnotes rather than a full-page list, and there can be discrepancies between the content of LLM-generated responses and the sources they cite~(\cite{liu2023evaluating}).

How will the differences between traditional and LLM-based search affect people's every day decision making?
LLM-based search could offer substantial benefits, providing an easier-to-use interface that speeds up complex tasks to help people accomplish their goals more quickly or free up time for them to aquire more information.
At the same time, fabrications in LLM-generated results could mislead people, and so while they might complete tasks more quickly, they might also make sub-optimal decisions based on inaccurate information.

In what follows, via a large randomized experiment, we empirically test how LLM-based search tools affect decision making and propose  and test interventions to mitigate overreliance on erroneous LLM-based responses.
Participants in our experiments were asked to solve a series of decision tasks that involved researching and comparing different products, and were randomly assigned to do so with either an LLM-based search tool or a traditional search engine.
In our first experiment, we find that participants using the LLM-based tool were able to complete their tasks more quickly, using fewer but more complex queries than those who used traditional search.
Moreover, these participants reported a more satisfying experience with the LLM-based search tool. 
When the information presented by the LLM was reliable, participants using the tool made decisions with a comparable level of accuracy to those using traditional search, however we observed overreliance on incorrect information when the LLM erred. 
Our second experiment further investigated this issue by randomly assigning some users to see a simple color-coded highlighting scheme to alert them to
potentially incorrect or misleading information in the LLM responses.
Overall we find that this confidence-based highlighting substantially
increases the rate at which users spot incorrect information, improving
the accuracy of their overall decisions while leaving most other
measures unaffected.
Together these results suggest that LLM-based information retrieval tools have promise for increasing the productivity of people engaged in decision tasks, and highlights the opportunity of communicating uncertainty to help people know when to scrutinize or further verify LLM output.

\section{Related Work}\label{sec:related-work}

In this work we measure how people make consumer decisions using a novel search tool that, holding all else constant, allows random assignment to traditional search or LLM-based search. While LLM-based search is a very recent innovation, this work has connections to past research the effects of generative AI on knowledge work and the rich literature on how people use search engines.

\cite{noy_experimental_2023} conducted an online experiment to evaluate
the impact of LLM-based writing assistants on worker productivity and
associated measures. The participants were assigned tasks that simulated
real work activities, such as writing press releases, brief reports, and
emails. Experienced professionals in corresponding fields evaluated the
participants' work and found that the AI assistant improved productivity
and enhanced the quality of writing in several
ways.

\cite{brynjolfsson_generative_2023} explored the influence of generative
AI on productivity in the customer service sector by investigating the
deployment of a GPT-based chat assistant. They discovered that it
positively impacted productivity, especially for lower-skilled workers,
and resulted in other beneficial outcomes (e.g., fewer
escalations). \cite{dell2023navigating} executed a controlled experiment with consultants at Boston Consulting Group, where they showed increased productivity for certain types of tasks, with more gains for lower skilled workers. Conversely, they also showed accuracy drops for tasks for which the LLM-based tools were known to have difficulty.

In the domain of software developer productivity, \cite{peng_impact_2023} conducted a controlled experiment using an LLM-based coding tool
(GitHub Copilot) to assess its impact on productivity. Developers who
were assigned coding tasks and randomly provided with LLM assistance
completed the tasks in less than half the time it took the control
group. The study revealed that certain groups (e.g., less experienced
developers) reaped more benefits. Earlier research by \cite{ziegler_productivity_2022} suggested that the rate of acceptance of code suggestions, rather
than their actual persistence in the final code, predicts developers'
perceptions of productivity. The findings of Peng et al. illustrate that
productivity improvements can lead to significant time
savings. In a field experiment, \cite{van2023algorithmic} shows the productivity gains in writing resumes and choosing candidates, as the use of LLM-based tools made cleaner resumes, which could have eliminated source of identification for employers, but instead made their choices more efficient.

Furthermore, our work investigates the potential overreliance
on LLM-based tools when their responses contain errors or fabrications,
a phenomenon that has been well documented (\cite{liu_evaluating_2023}). Although
there has been significant effort to algorithmically identify these
mistakes or generate calibrated probabilities for response correctness,
most of it has focused on relatively straightforward scenarios, such as
standardized tests or similar banks of question-answer pairs (\cite{kadavath_language_2022,lin_teaching_2022,yin_large_2023}). Recent research has
shown promise for more complex, real-world scenarios with the use of
color-coded highlighting of LLM-generated code to help programmers
direct their attention to potentially problematic output (\cite{vasconcelos_generation_2023}). In our study, we adapt this approach to LLM-based search,
using color-coded highlighting to alert users to potentially misleading
information in LLM-generated
responses.

This paper builds on a comprehensive literature on
how people use search and adapt their search behavior as the technology
evolves, where LLMs are the latest of many disruptions.  \cite{bates_design_1989} introduced models of online search, importantly
focusing not merely on what people were beginning to do, but also how
the skills of users co-evolved with the interface. \cite{bennett_modeling_2012}
offered a detailed categorization of elements of the search session---from terms to the number of queries---that we expand upon in this paper. \cite{liu_deconstructing_2021} surveyed a broad spectrum of research into what users aim to
achieve in various search sessions. Numerous studies have examined
changes in search behavior, from the impact of new interfaces (\cite{bates_design_1989}) to auto-completion features (\cite{mitra_user_2014}). This research
also builds upon the literature about the conversion journey and how
people use search engines to find products (\cite{ramos_search_2008}).

Our research builds upon the existing literature in several ways.
First, in contrast to studying writing or coding, we focus on how LLMs impact search and information retrieval.
Second, we focus on the domain of consumer decision making, a broad category that, to the best of our knowledge, has not yet been explored in the literature on LLM-assisted productivity.
And third, we explore solutions to mitigate overreliance in situations where LLMs return unreliable information.

\section{Domain and Research
Questions}\label{sec:domain-and-research-questions}

The domain we investigate in this work is product research by consumers.  
Search engines play a vital role in what is called the conversion
journey for goods and services, that is, the process by which consumers
move from interest in a product category to a consideration set, a final
product purchase, and even post-purchase activities like seeking support
or choosing accessories. In a simple model, consumers begin with an
interest in a product category and use search engines to gather
preliminary information. They navigate in and out of the search engine
as they explore different products. Those who opt to continue the
journey move closer and closer to points of purchase, often found through
search engines. Because of its integral connection to conversion
journeys, search advertising has grown into a substantial industry,
having revenues of \$84 billion in 2022 and making up the largest
revenue share among all internet advertising forms (\cite{iab_pwc_2023}).

\begin{figure}[t]
  \centering
  \includegraphics[width=\linewidth]{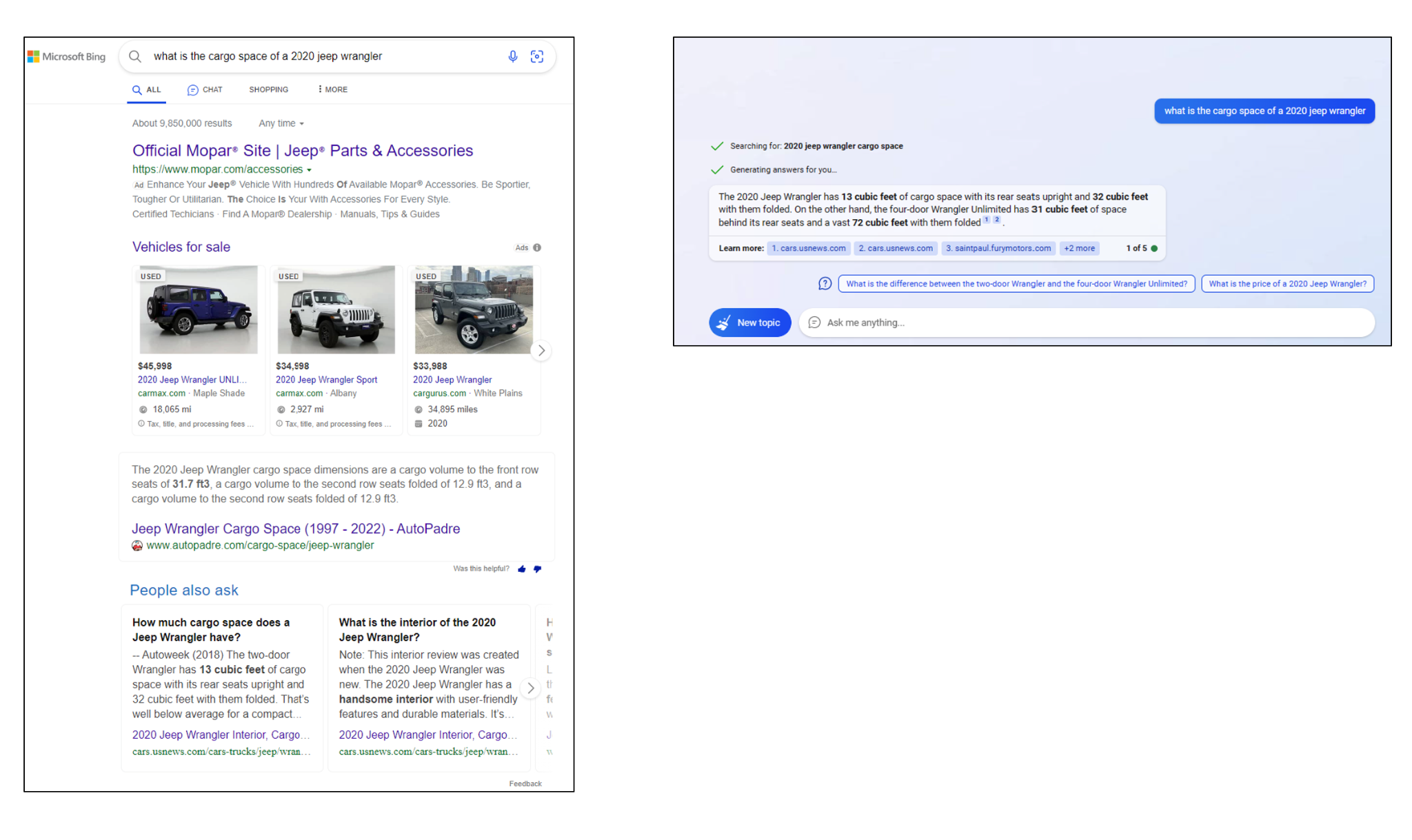}
  \caption{Example of the same query “what is the cargo space of a 2020 jeep wrangler” in (left) Bing’s traditional search on May 15, 2023 and (right) Bing’s conversational search on May 15, 2023. \label{fig:queries}}
\end{figure}

In particular, we focus on researching the purchase of a Sports Utility
Vehicle (SUV). Figure \ref{fig:queries} provides screenshots of a traditional search
engine results page and how a currently-running LLM-based search tool, Bing Chat,
responds to queries about product dimensions. In both panels the query
was ``what is the cargo space of a 2020 jeep wrangler?'' In the left
panel, Bing's traditional search provides a mixture of advertisements, organic results that link to relevant webpages, and highlighted ``instant answers'' that come from a snippet of
a linked website. In the right panel, using Bing's chat-based
search, conversational instant answers appear in natural language, with detailed information generated via an LLM-based summary of relevant webpages. 

Suppose someone was in the market
for an SUV that offered ample cargo space (to transport packages in) but
with minimum total length (to facilitate parking). How might they go
about searching for a vehicle that maximizes this ratio of cargo space
to total length? Looking at data from a traditional search engine (Table
\ref{tab:products}), we see that people often search for one vehicle at a time and
information on one dimension (e.g., cargo space) at a time. Perhaps our
hypothetical searcher for a high volume, short length SUV might issue a
series of simple queries when using a traditional search engine. A
series of 1 product and 1 dimension searches could, for example, be: ``jeep wrangler
total cargo space'' in one query and ``jeep wrangler length'' in
another, followed by some calculations to figure out the ratio of cargo space to length.

How might the search experience differ with LLM-enabled
search tools? 
In this environment, a user might instead simply issue a complex, natural language query that directly addresses the decision they are looking to make.
For instance, when
choosing between two SUVs, a person could issue a complex query like
``Which vehicle has a larger cargo space to length ratio, a Jeep Wrangler or a Hyundai Santa Fe?''
That said, two key open questions remain: (how quickly) will
users adapt to this new style of search, and what happens if the LLM response
contains mistakes or fabrications?

To gain insight into these questions, we conducted two online experiments in
which participants were randomly assigned to complete a consumer product
research task using either an LLM-based search tool or a traditional
search tool. We designed these experiments to focus on the following questions:

\begin{itemize}
\item
  \textbf{Research question 1 (efficiency):} How will task completion
  time and the number (and complexity) of queries issued differ between
  participants in the LLM search condition as opposed to the traditional
  search condition?
\item
  \textbf{Research question 2 (accuracy):} How will the accuracy of
  decisions differ between participants in the LLM search condition as
  opposed to the traditional search condition?
\item
  \textbf{Research question 3 (perceptions):} How will the user
  experience and perceived reliability of results differ between
  participants in the LLM and traditional search conditions?
\item
  \textbf{Research question 4 (confidence and errors):} How will
  participants handle mistakes in the LLM responses with and without
  cues about the confidence reported information?

\end{itemize}

\section{Experiment 1}\label{experiment-1}

\begin{figure*}[h!]
 \captionsetup[subfigure]{belowskip=2pt}
\centering
\begin{subfigure}[b]{0.9\textwidth}
        \fbox{\includegraphics[width=1\linewidth]{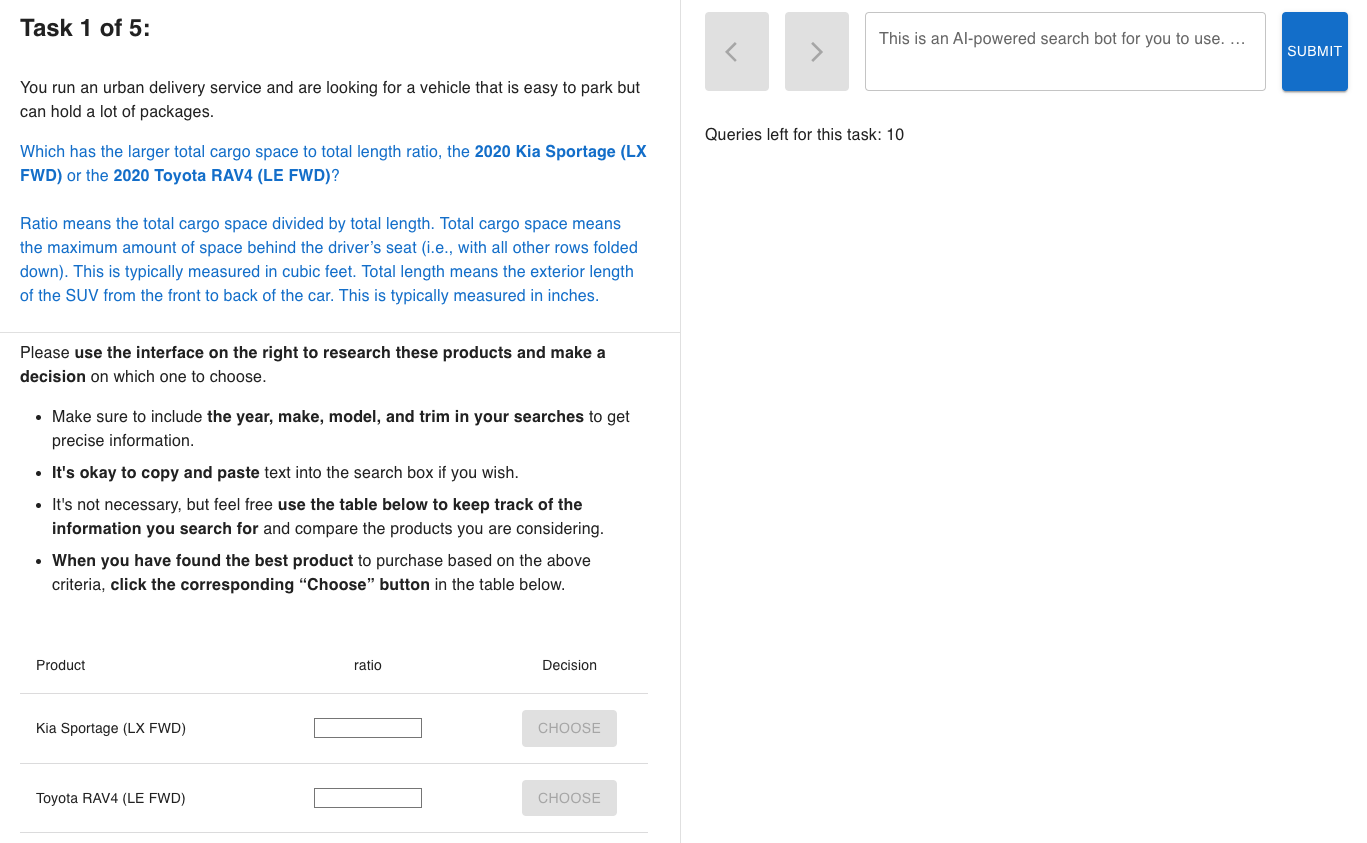}}
        \caption{The main task interface. Participants are asked to choose between two vehicles. Instructions on the scenario and the metric of interest to make a decision are provided on the left, while the search tool is on the right. A notepad on the bottom left is also available for keeping track of found information. \label{fig:interface}}
\end{subfigure}

\begin{subfigure}[b]{0.9\textwidth}
        \includegraphics[width=1\linewidth]{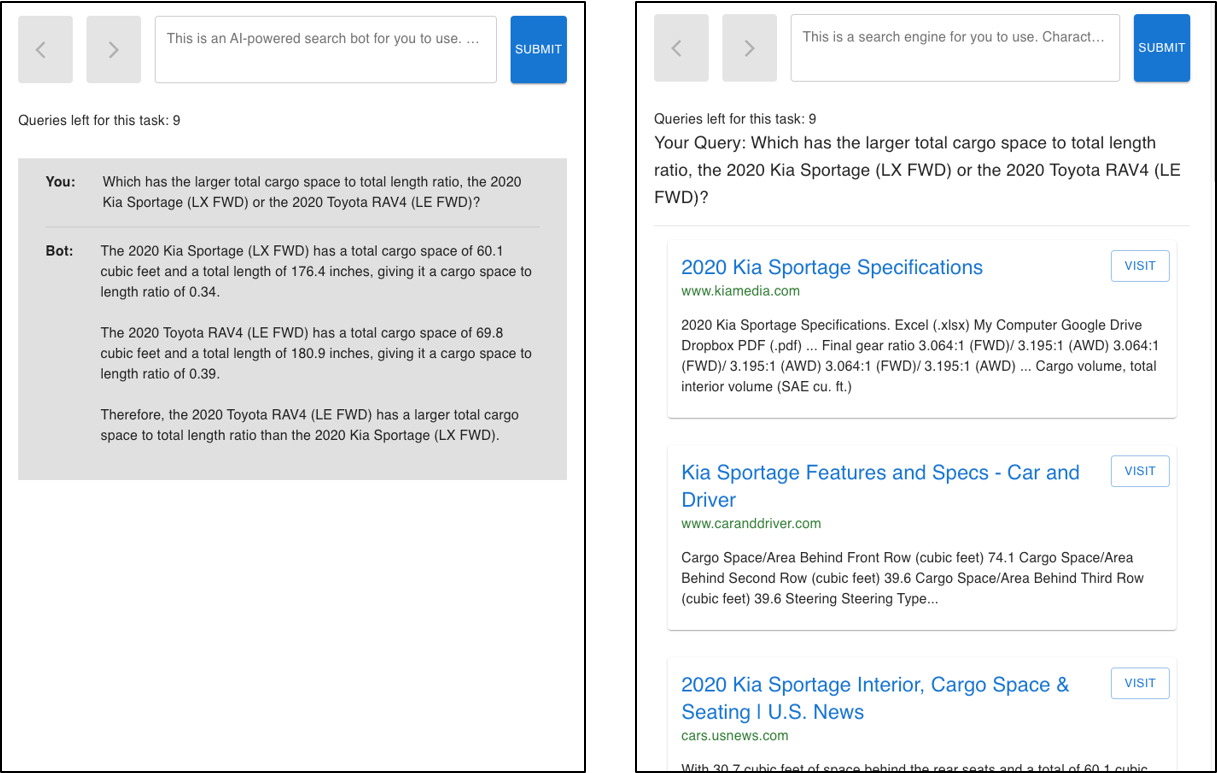}
        \caption{Search tool response interface for both conditions in this experiment: the experimental AI-powered search tool (left) and the experimental search engine (right).\label{fig:tool_resp}}
\end{subfigure}
\hfill
\caption{Screenshots of the interface for Experiment 1.}
\end{figure*}

We designed a task in which participants assume the role of running an
urban delivery service (Figure \ref{fig:interface}). In this role, they are looking to
purchase a vehicle to meet their business needs and are choosing between
two SUVs. To capture common criteria for choosing a vehicle for
deliveries (ability to hold many packages and parking flexibility), we
defined the main metric for choosing a vehicle as the cargo space to
total length ratio. Cargo space in this case is defined as the maximum
amount of space behind the driver's seat, with all other seats folded
down, and total length is defined as the exterior length of the vehicle.
Therefore, a higher cargo space to total length ratio would translate to
a vehicle that is better suited to meet the delivery service business
needs.
This design ensures that there is both a correct answer to each task participants are given, and provides them with clear criteria to be used in making their decisions.

Participants were invited to complete a series of five of the above
defined tasks, where the goal for each task was to determine the best
option from a randomly generated vehicle pair. We varied the type of
assistance they received for their search in making their decisions in a
two condition, between-subjects design. In one condition, participants
were provided with an experimental search engine built using the Bing
API. Similar to traditional search, the experimental search engine
returned a series of clickable links with descriptions based on the
input query, that participants would be able to visit to get more
information. In the second condition, participants had access to an
experimental LLM-based search tool, built using GPT 3.5.\footnote{At the time these experiments were conducted, and at the time of writing, Bing did not have an API for interacting with their LLM-based search tool, Bing Chat.}. The LLM-based
tool responded in natural language to participant queries and had no
conversational capabilities to provide a tight control against the traditional search condition. No information on the technology used
behind either search tool was provided to participants, but in both
conditions participants were given a quick tutorial on how to use the
corresponding tool and what to expect from it, as shown in Appendix I.
Figure \ref{fig:tool_resp} shows the difference in responses from the search tool for the
two conditions.

We imposed a limit of 10 searches per task and a limit of 1,000
characters per search. In addition, participants had to complete at
least one search to be allowed to make a decision and thus proceed to
the next task. For both conditions, participants had access to their
full search history and could revisit search tool responses at any time.
After completing all available tasks, participants were also asked to
complete a brief survey about their experience to conclude the
experiment.

The LLM-based search tool was issued the following pre-prompt in order
to provide a consistent experience for participants and to signal to
participants that it is not a conversational tool: ``You are a search
engine to be used for finding facts about motor vehicles and doing math.
If you are given a query about the features of a commercial car, truck,
or SUV, do your best to answer it. If you are given a query that
involves doing math, do your best to answer it. If you are given a query
that seems like it's trying to refer to a previous conversation, respond
with 'Sorry, I do not have the ability to refer to information from past
questions or answers.' Otherwise respond with 'Sorry, that does not seem
like a relevant query. Please try again.' Show your work.'' This
pre-prompt was not visible to participants.

In order to increase the chances that participants in the LLM-based tool
condition would see a case on which the LLM-based tool reports erroneous
information, we set the last item to be identical for all participants
and to involve a vehicle with which the LLM-based tool tends to report
the incorrect amount of cargo space (the 2020 Toyota 4Runner). In this
particular case the LLM-based tool is prone to confuse the cargo space
with the seats up for the cargo space with the seats down, thus
reporting that the SUV with the largest cargo space (with seats down, as
specified in the instructions) actually has the smallest cargo space.
Participants in both conditions saw this item, in the same position
(task 5).

We recruited 90 U.S. based participants from Amazon Mechanical Turk. For
qualifications, we required at least 2500 HITs approved with a 99\%
minimum approval rate, along with an additional Masters system
qualification. Participants were paid \$4 for completing the experiment,
with no performance bonuses.

\subsection{Results for Experiment 1}\label{results-for-experiment-1}

\begin{figure*}[h!]
 \captionsetup[subfigure]{belowskip=2pt}
\centering
\begin{subfigure}[c]{0.90\linewidth}
\includegraphics[width=0.9\linewidth]{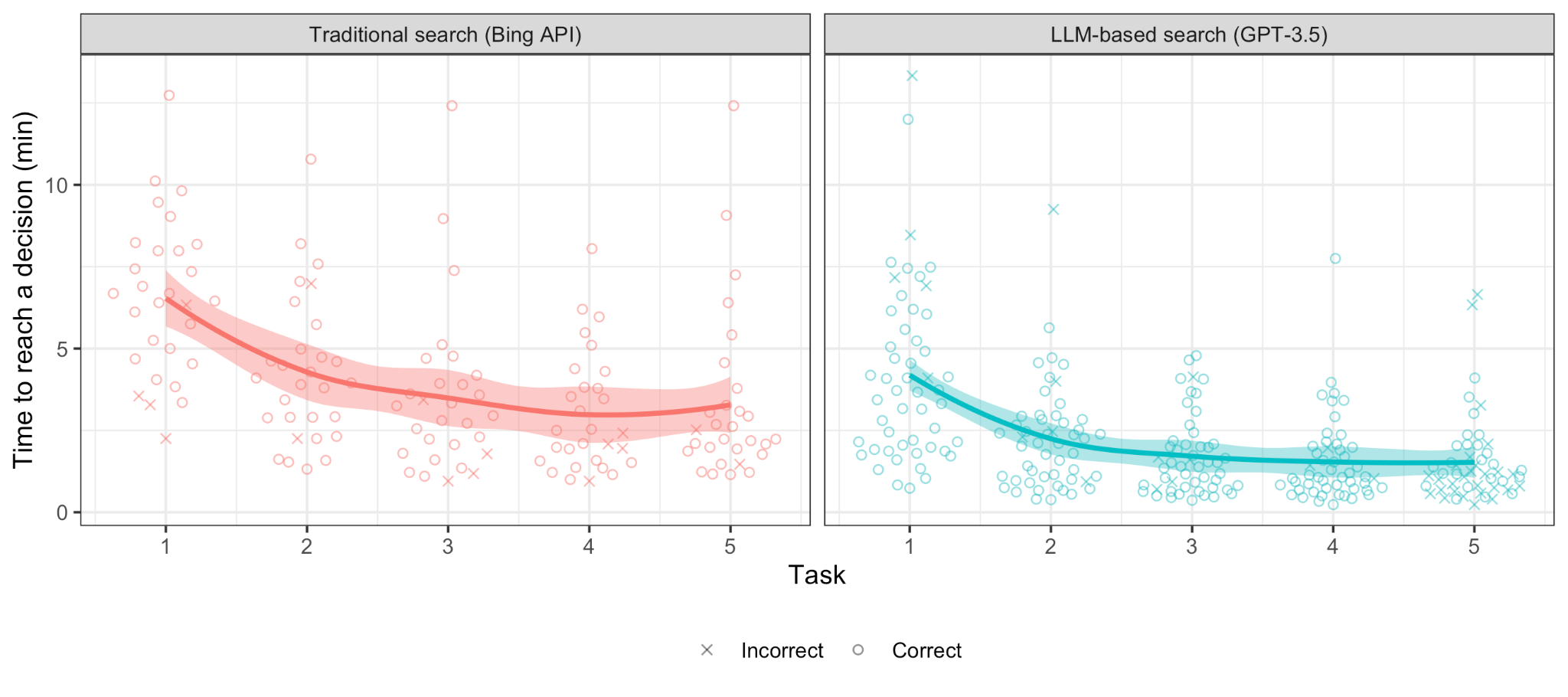}
  \caption{Time to reach a decision by condition and
task. Participants answered questions about five pairs of vehicles,
which each question counting as one task (horizontal axis). Each point
represents one participant's time taken for the task. \label{fig:res_time} }
\end{subfigure}

\begin{subfigure}[c]{0.90\linewidth}
\includegraphics[width=0.9\linewidth]{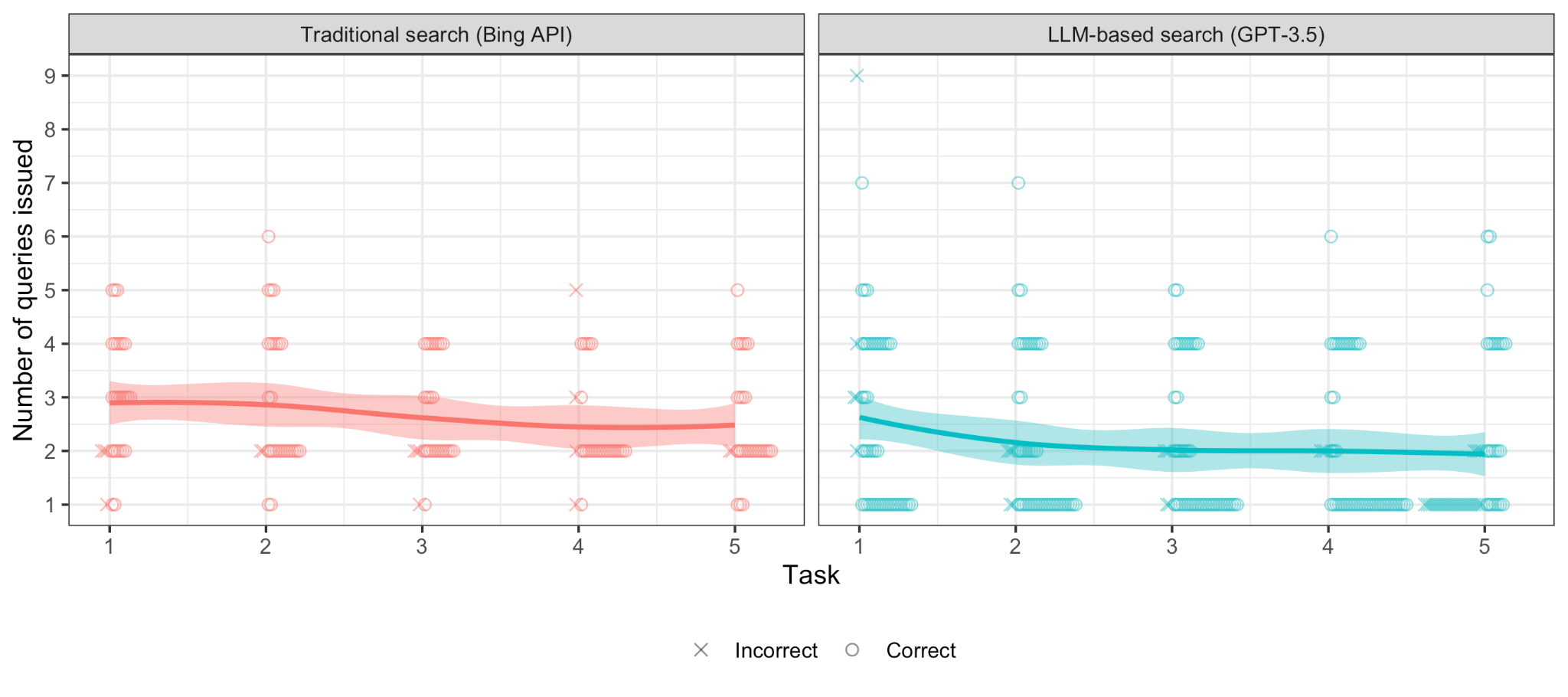}
        \caption{Number of queries issued by condition and task. Each point
represents one participant's number of queries for the task. \label{fig:res_queries} }
\end{subfigure}
\hfill
\caption{Experiment 1: Efficiency results}
\end{figure*}

\paragraph{Efficiency.} As shown in Figure \ref{fig:res_time},
participants took less time to complete the task in the LLM-based search
condition relative to the traditional search condition, a pattern which
is apparent as early as the first question. In both conditions we see a
learning effect where participants are slower in the first task compared
to subsequent tasks. Participants are simultaneously learning about the
task and the domain, while also learning about the functionality of the tool they are
using. In addition to the time to respond being less in the LLM-based
search condition, the variance was also
lower.

A linear mixed model fit to task duration confirms this. Specifically,
we modeled the log task time based on a random effect by participant id,
controls for task number, and a fixed effect for condition (\texttt{lmer:
log10(task\_duration\_full) \textasciitilde{} (1\textbar worker\_id) +
as.factor(task\_num) + condition}). The fixed effects estimates revealed
statistically and practically significant effects of task number and
condition on the log-transformed task duration. The LLM-based condition
significantly reduced the log-transformed task duration compared to the
traditional search condition (Estimate = -0.31613, SE = 0.05542, t(78) =
-5.70, p \textless{} .001), and all tasks were faster, on average,
relative to the first across conditions. The estimated average task
durations, back-transformed from the log10 scale, were 3.4 minutes (95\%
CI {[}2.8, 4.1{]}) for the traditional search condition and 1.6 minutes
(95\% CI {[}1.4, 1.9{]}) for the LLM-based search condition, a roughly
50\% reduction for the LLM-based tool.

\begin{figure}[t]
  \centering
  \includegraphics[width=\linewidth]{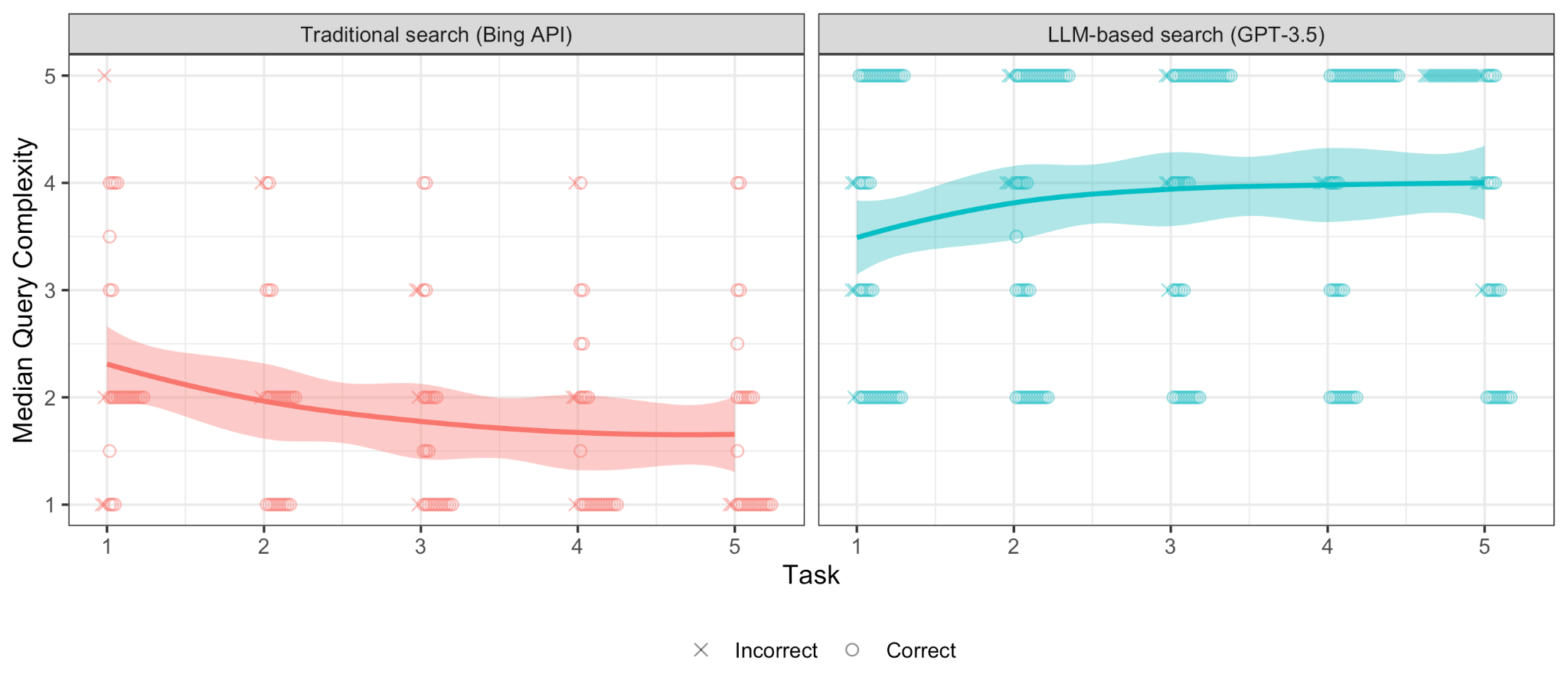}
  \caption{Complexity of queries issued by condition and
task (Experiment 1). Each point represents an average of the complexity of all of the
queries issued by a given participant in a given task. \label{fig:res_complexity} }
\end{figure}

Consistent with participants taking less time to answer with the
LLM-based tool, participants issued fewer queries with the LLM-based
tool as well, as shown in Figure \ref{fig:res_queries}. With the LLM-based tool, most
participants issued one query in all the tasks, while with the search
tool, two queries was the most common pattern. Interestingly, many
participants in the traditional search condition navigated to product
information or comparison pages that allowed them to get both
measurements for both vehicles in fewer than four, simpler queries for
one product and dimension at a time.

We tested this difference with a generalized linear model, using a
Poisson link function to model the number of queries by participant id
as a random effect, task number as a control, and condition as a fixed
effect (\texttt{glmer: num\_queries \textasciitilde{} (1\textbar worker\_id) +
as.factor(task\_num) + condition, family = poisson, \\ data = task\_data}).
The model revealed a modest but statistically significant main effect of
condition on the number of searches (Estimate = -0.26, SE = 0.12, z =
-2.244, p = 0.02). The estimated average number of queries made was 2.5
(95\% CI {[}2.1, 3.0{]}) for the traditional search condition and 1.9
(95\% CI {[}1.7, 2.2{]}) for the LLM-based search condition.

While participants take less time and issue fewer queries in the
LLM-based search condition, they make up for fewer queries by asking
more complex queries. We average the complexity of each person by task
in Figure \ref{fig:res_complexity}, where complexity is a number between 1 and 5 representing
the number of unique elements of interest noted in the query. This could
include 0, 1, or 2 products, 0, 1, or 2 dimensions, and 0 or 1 math
question for the ratio of cargo space to length. Traditional search
starts up above 2 on average and goes down, while LLM-based search
starts near 3 on average and goes up. Most of the gains, in both
conditions, are between the 1\textsuperscript{st} and
3\textsuperscript{rd} task. Most LLM-based searches are either 2 or 5,
with comparatively few at 3 or 4. Similar to the telemetry data noted in
Table \ref{tab:products}, a surprising amount of traditional search is just 1 (i.e., a
single product and no dimensions). While it takes a lot of time and
queries, these searchers almost always make correct final decisions.

We tested this difference with a generalized linear model, using a
Poisson link function to model the complexity of queries by participant
id as a random effect, task number as a control, and condition as a
fixed effect (\texttt{glmer: complexity \textasciitilde{} (1\textbar worker\_id)
+ as.factor(task\_num) + condition, family = poisson, data =
task\_data}). The model revealed a statistically significant main effect
of condition on the complexity of queries (Estimate = 0.65, SE = 0.09, z
= 7.38, p \textless{} 0.001). The estimated average complexity of
queries made was 1.8 (95\% CI {[}1.6, 2.1{]}) for the traditional search
condition and 3.4 (95\% CI {[}3.1,3.8{]}) for the LLM-based search
condition.

\begin{figure}[t]
  \centering
  \includegraphics[width=0.60\linewidth]{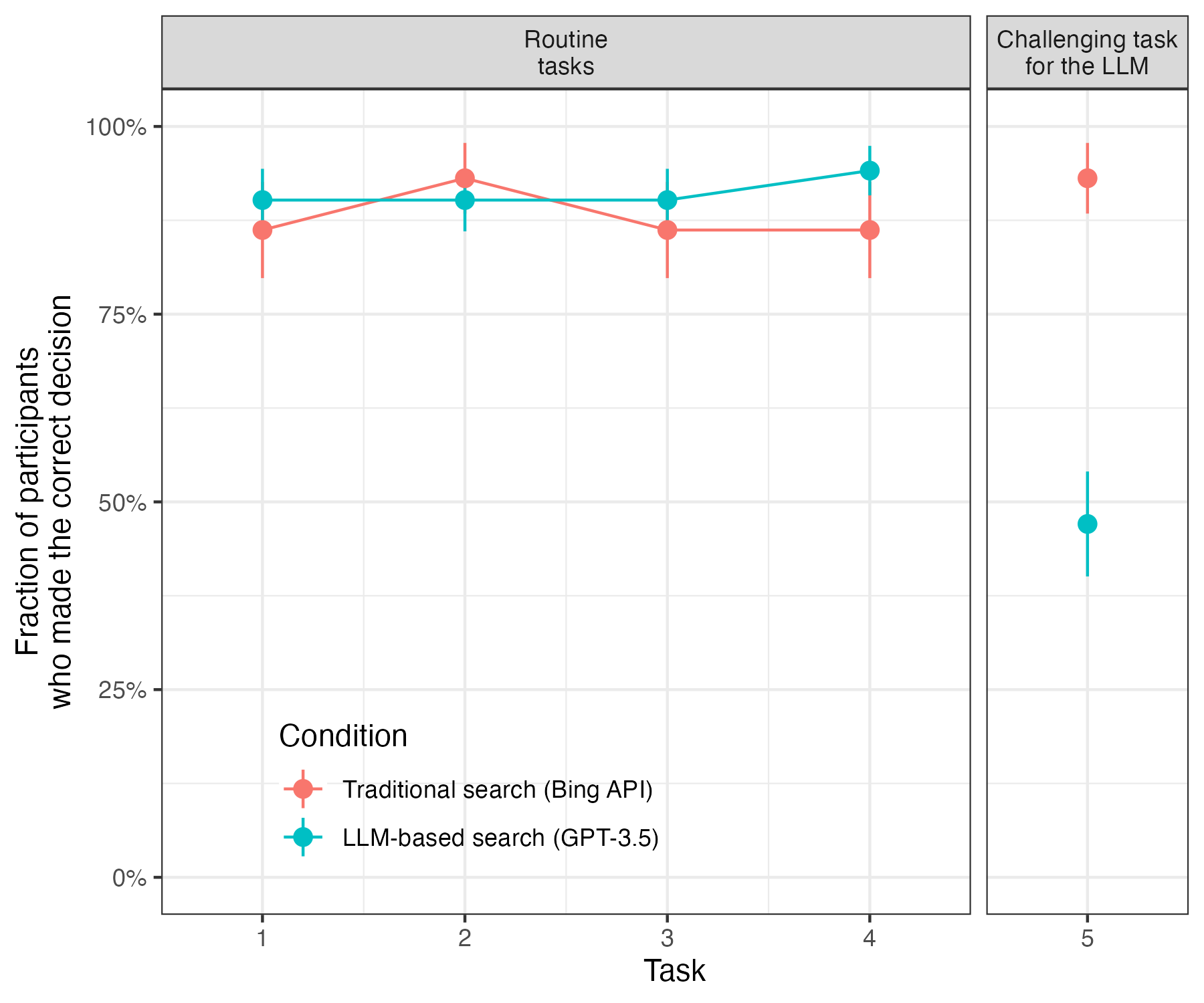}
  \caption{Accuracy by condition (Experiment 1). The first four tasks
are routine (comparisons between 8 popular SUV models), whereas the
fifth is a comparison selected for which the LLM tends to err. Points
represent means and error bars are plus or minus one standard error. \label{fig:res_accuracy} }
\end{figure}

\paragraph{Accuracy.} Figure \ref{fig:res_accuracy} shows accuracy by
task. For the routine tasks (comparisons between 8 popular,
randomly-paired SUV models) accuracy was comparable between the two
conditions, despite the traditional search users spending more time and
issuing more queries to answer the questions. On the task designed to be
difficult (i. e., one where the LLM tends to err), accuracy drops
greatly due to mistakes in the LLM's responses, largely due to it
returning the cargo space with the second row of seats up instead of
down.

To compare accuracy between conditions we fit a generalized linear model
for the first four ``routine'' tasks. Specifically, we modeled whether
participants made the right choice for each task, accounting for random
effects by participant id, controls for the task, and a fixed effect for
the condition they were assigned to (traditional vs. LLM-based search),
with a logistic model (\texttt{glmer: is\_correct\_decision \textasciitilde{}
(1\textbar worker\_id) + as.factor(task\_num) + condition, family =
binomial}). The fixed effects estimates revealed no significant effect of
condition on the likelihood of making a correct decision for routine
tasks (z = 0.99, p = 0.33). The estimated probabilities of making a
correct decision, averaged over routine tasks, were 92.3\% (95\% CI
{[}83\%, 97\%{]}) for the traditional search condition and 95.3\% (95\%
CI {[}89\%, 98\%{]}) for the LLM-based search condition.

We used a separate generalized linear model to investigate accuracy in
the final task, which was constructed to be challenging for the LLM.
Specifically, we fit a logistic model with a fixed effect for the
condition to predict whether participants made the right choice for this
task (glmer: is\_correct\_decision \textasciitilde{} condition, family =
binomial). The fixed effect estimate showed that the LLM-based search
condition had a significant negative effect on the likelihood of making
a correct decision compared to the traditional search condition
(Estimate = -2.72, SE = 0.79, z = -3.46, p \textless{} .001). The
estimated probabilities of making a correct decision were 93\% (SE =
5\%, 95\% CI {[}76\%, 98\%{]}) for the traditional search condition and
47\% (SE = 7\%, 95\% CI {[}34\%, 61\%{]}) for the LLM-based search
condition.

The previous figures hint at what happens in the final task:
participants that have very complex queries are much more likely to get
the task wrong. We dissected the query stream of the 51 participants in
the LLM-based search condition: 30 of them did just one query in the
final task with 23 getting the wrong answer and 7 getting the right
answer. All 23 were given the wrong answer from their query, while all 7
that got the right got the correct answer from their query (most of the
23 that got wrong answers cut and paste the directions into the query,
while the 7 that got right answers wrote in variations of the
directions). Ten respondents had 2 queries (6 were correct and 4 wrong),
and again their accuracy was directly driven by the accuracy of the
answer to their queries. Eleven participants issued 4 or more queries:
all of their queries returned correct answers and they all picked the
correct option. No participant re-queried a product and dimension after
seeing a wrong answer. Again these wrong answers were always driven by
the LLM-based output giving a seats-up cargo space, rather than
seats-down, giving a very small value for one particular SUV: yet, no
participant issued a further query after getting an abnormally small
cargo space reading on the fifth task.

\begin{figure}[t]
  \centering
  \includegraphics[width=0.75\linewidth]{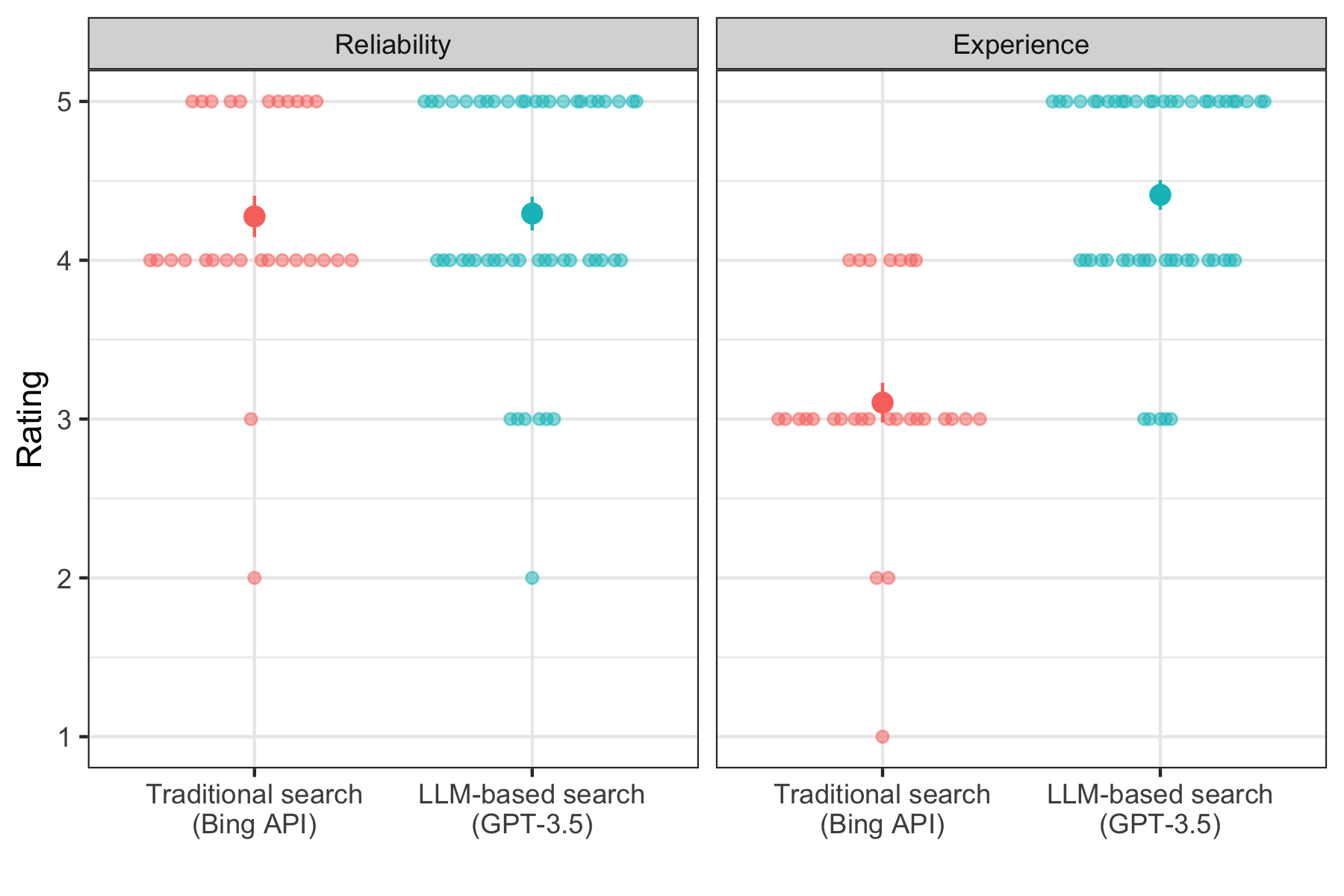}
  \caption{Results on user perceptions (Experiment 1). Each smaller
point represents one participant's response, the larger points show the
mean by condition with error bars of +/- 1 SE.
 \label{fig:res_user} }
\end{figure}

\paragraph{User experience and perceived reliability.} In the survey
at the end of the experiment we asked participants to rate the overall
search experience they were shown and to rate the reliability of the
results they were shown, both on 5 point Likert scales, with 1 being the
worst and 5 being the best. As seen in Figure \ref{fig:res_user}, perceived reliability
was similar between conditions and overall quite high, which is
remarkable in that many using the LLM-based tool were exposed to an
erroneous response on the last task (by design). We find no
statistically significant difference in participants' subjective ratings
of the reliability of the results they were shown (t(62.03) = 0.11, p =
0.91), indicating that users in the LLM condition who saw unreliable
information were unaware of the errors made in the LLM output. Where
experience is concerned, users strongly prefered completing the task
with the LLM-based tool compared to traditional search. Overall we find
that participants strongly preferred the LLM-based search experience
(with average rating of 4.41) compared to the traditional search
experience (with an average rating of 3.10), a statistically significant
difference (t(58.00) = 8.38, p \textless{} .001).

\section{Experiment 2}\label{experiment-2}

In the previous experiment we saw that while LLM-based search helped
participants arrive at decisions faster than traditional search, these
decisions were often---but not always---of the same quality.
Specifically, when the LLM response contained inaccurate information, it
was difficult for participants to spot these mistakes due to a lack of
cues about the veracity of the information they were shown. We designed
our second experiment to investigate how people react to explicit cues
that convey confidence in the responses generated by the LLM, and how
this affects their decision making.

\begin{figure}[t]
  \centering
  \includegraphics[width=\linewidth]{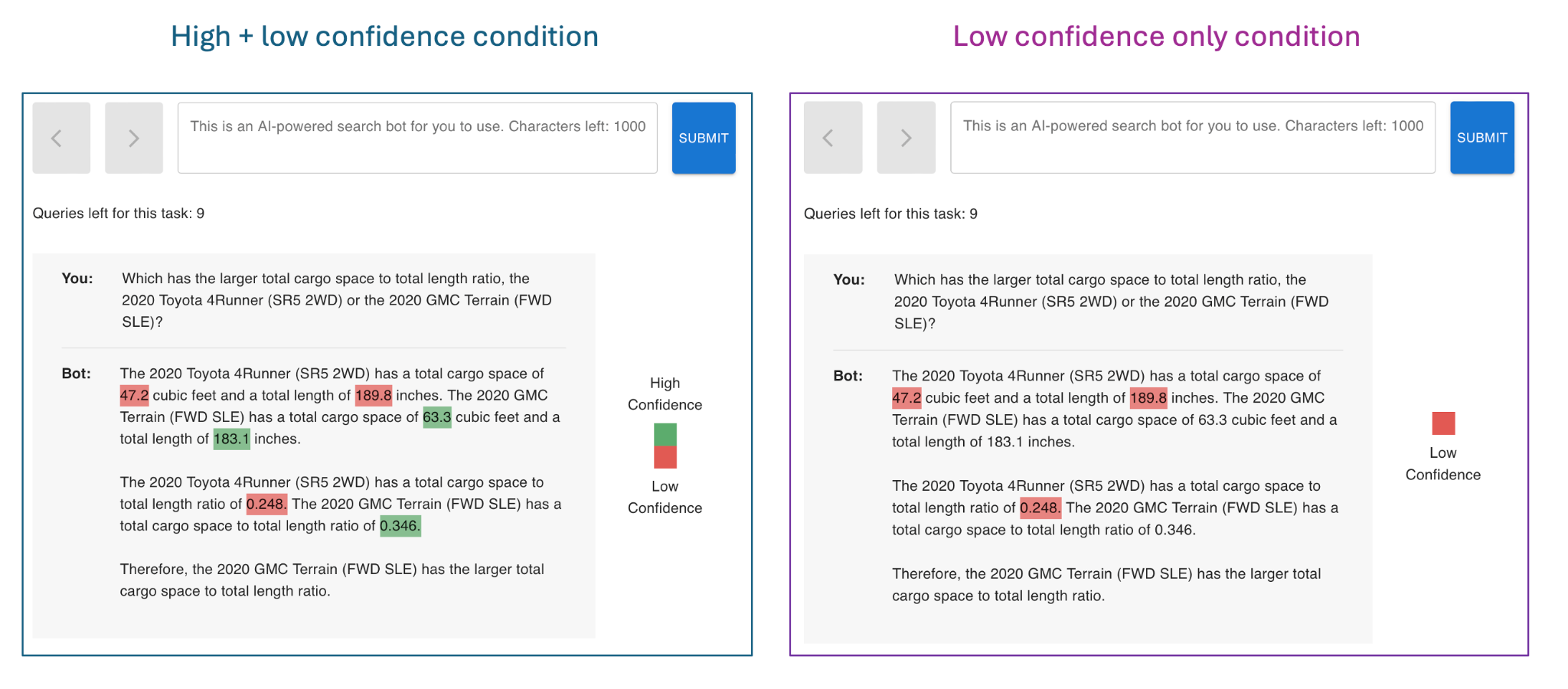}
  \caption{The two treatments tested in Experiment 2:
highlighting of both low and high confidence measurements (left) and
only low confidence measurements (right). There was an additional
control condition in which no highlighting was shown, mirroring
Experiment 1.
 \label{fig:interface_st_2} }
\end{figure}

This experiment was a three condition, between-subjects design where all
participants were assigned to use the same LLM-based search tool that
generated the same responses to a given query.\footnote{We used GPT-3 in
  this experiment in order to have access to token probabilities, which
  were not available with GPT-3.5 used in the first experiment.} The only
thing that varied between conditions was how the numerical measurements
in the responses were displayed visually via color coding. In the
control condition, participants saw answers similar to those shown in
Experiment 1---plain text without any cues about the veracity of
measurements in the response. In each of the two treatment conditions,
participants saw confidence-based color highlighting for numerical
measurements contained in responses. As depicted in Figure \ref{fig:interface_st_2}, the ``High
+ low confidence'' condition showed green highlighting for ``high
confidence'' measurements and red for ``low confidence'' ones, whereas
the ``Low confidence only'' condition showed red highlighting for ``low
confidence'' measurements only. The highlighting of each measurement was
based on the token generation probabilities provided by GPT-3, with a
generation probability of less than or equal to 50\% displayed as a red
highlight and greater than 50\% displayed as a green
highlight.\footnote{Specifically, for measurements greater than 1, we
  used the token probability for the whole number token only (to the
  left of the decimal), whereas for measurements less than 1 we used the
  token probability for the decimal token only (to the right of the
  decimal. For example, for ``47.2'' the token probability for ``47'' is
  used, whereas for ``0.248'' the token probability for ``248'' is used.}

The procedure was nearly identical to Experiment 1. Participants
completed a sequence of three decision tasks comparing pairs of SUVs on
the same criteria as in our first experiment (the ratio of total cargo
space to total length). And, as in the first experiment, all but the
last task were ``routine'' for the LLM in that there was a high
likelihood of it returning correct information with high confidence,
whereas the third task was once again ``challenging'' for the LLM and
likely to contain inaccurate information, but with low confidence. We
achieved this by pre-prompting the LLM with ground truth measurements
for the vehicles involved in each task \emph{on everything except the
first query of the third task}. This meant that the first and second
tasks largely returned accurate information with high confidence, but
the first query of the third task often contained mistakes that were
highlighted as low confidence. So if participants issued queries of the
form ``Which has the larger total cargo space to length ratio the 2020
Toyota 4Runner or the the 2020 GMC Terrain'' on the first query of the
third task, those in the treatment condition would see cues about
potentially unreliable information in the LLM response. The key question
in this experiment was whether participants in the treatment conditions
would take note of these low confidence cues and issue subsequent
queries to double check the information they were shown.

We recruited 120 U.S. based participants from Amazon Mechanical Turk
from a vetted pool of high-effort workers. For qualifications, we
required at least 2,500 HITs approved with a 99\% minimum approval rate.
Participants were paid \$5 for completing the experiment, with no
performance bonuses.

\subsection{Results for Experiment 2}\label{results-for-experiment-2}

As in our first experiment, we analyzed efficiency, accuracy, and
perceived experience across all conditions, but in this experiment we
compare the three different treatments of confidence highlighting in
LLM-based search instead of contrasting LLM-based search with
traditional search.\footnote{Of note, results are not directly
  comparable between the two experiments because Experiment 1 used
  GPT-3.5 whereas Experiment 2 used GPT-3 due the need for token
  probabilities.} For brevity we include only top-level results on the
accuracy and perceived experience here, with the remaining results
presented in the Appendix.

 \begin{figure}[t]
  \centering
  \includegraphics[width=0.66\linewidth]{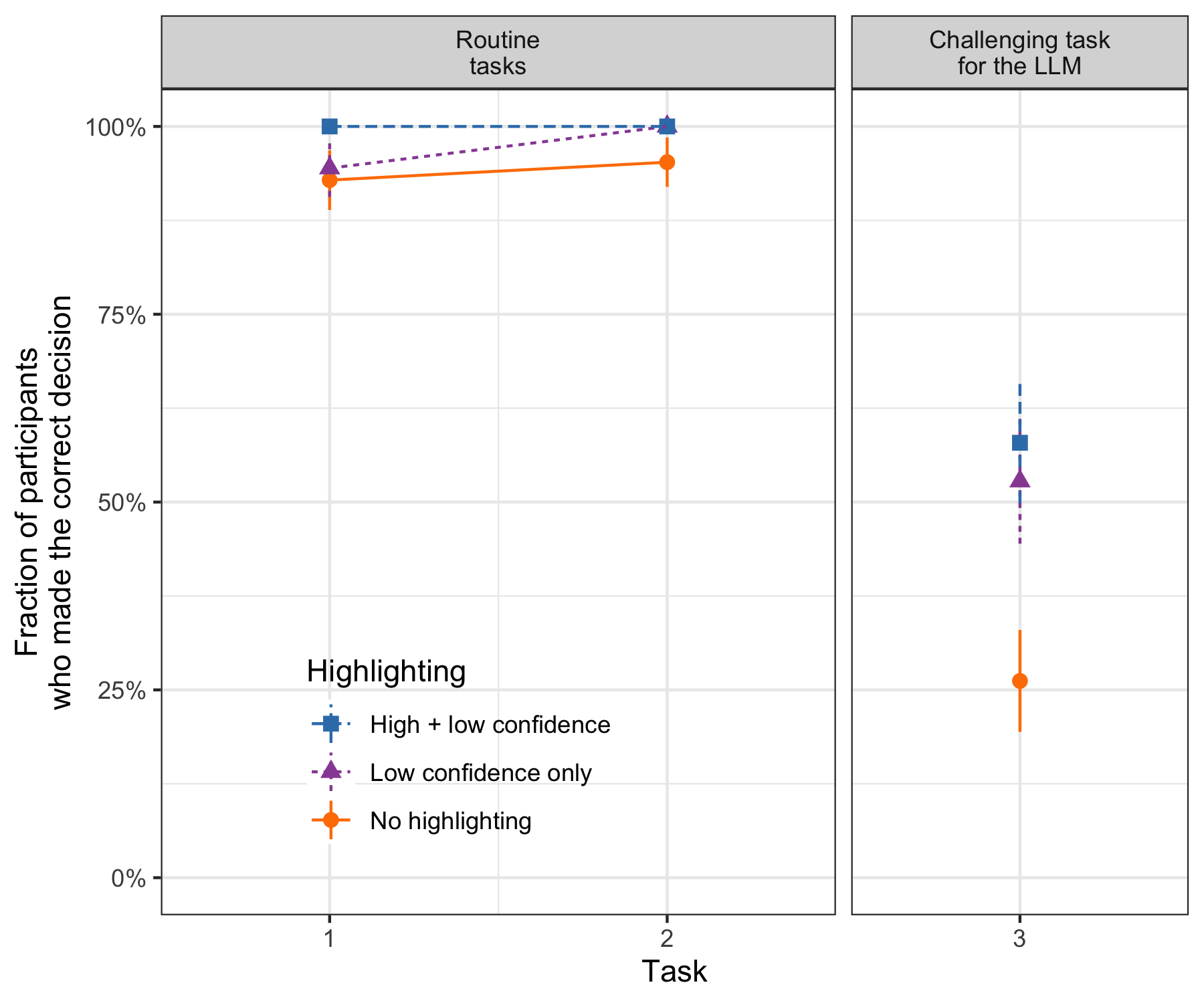}
  \caption{Accuracy by condition (Experiment 2). The first two tasks
are routine, whereas the third is a comparison selected for which the
LLM tends to err. Points represent means and error bars are plus or
minus one standard error.
 \label{fig:res_acc_st_2} }
\end{figure}

\paragraph{Accuracy.} As in our first experiment, for the routine tasks (tasks 1 and 2) where
the LLM provided largely reliable information with high confidence,
accuracy was comparable between all three conditions, and quite high
(Figure \ref{fig:res_acc_st_2}). However, for the challenging task (task 3) where the LLM
provided less reliable information on the first query, we see a dramatic
difference between conditions: while accuracy plummets to 26\% in the
control condition without any confidence highlighting, accuracy in each
of the treatment conditions was substantially higher---58\% for the high
+ low confidence condition (t(74.47) = -2.98, p \textless{} 0.01) and
53\% for the low confidence only condition (t(70.36) = -2.44, p = 0.02).
In this case, both showing high and low confidence cues and simply
flagging low confidence information more than doubled accuracy in the
decision task.

As shown in additional plots in the Appendix, the increased accuracy in
the treatment conditions is largely due to participants issuing their
initial query, seeing measurements flagged as low confidence, and
issuing follow-up queries to double check the information they were
shown. Whereas most participants in the control condition made a
decision after one query, the majority of participants in the treatment
conditions issued two or more queries, costing them some additional
time, but more often leading to the correct decision. To narrow in on
the most affected participants, 19 participants in the control
condition, in both tasks 2 and 3, provided a complete query for their
first query (i.e., a query that asked the comparison between both
vehicles on the ratio of both dimensions): in both tasks only 2 asked a
meaningful follow-up query despite all 19 getting the correct answer in
task 2 and wrong answer in task 3. In comparison, for participants in
the treatment conditions the amount issuing complete queries as the
first query jumped from 24 to 31 between tasks 2 and 3, but the amount
asking a meaningful follow-up query jumped from 5 to 15. So, the
treatment conditions had a slightly high rate of follow-up queries
(relative to the control condition) even when the answer was correct,
but had a much higher rate of follow-up queries when the answer was
incorrect.

Tying this back to experiment 1, the error the LLM-based search tool
created was generally similar: a very small cargo space was assigned to
a large SUV (due to outputting the seats up versus seats down measurement). In
experiment 1 (and the control condition in experiment 2) participants
were unlikely to issue a follow-up query when confronted with this small
cargo space alone, but in the treatment conditions in experiment 2
participants were increasingly likely to follow up when confronted with
both the small cargo space and a confidence-based color highlighting
indicating uncertainty in that datapoint.

\begin{figure}[t]
  \centering
  \includegraphics[width=0.75\linewidth]{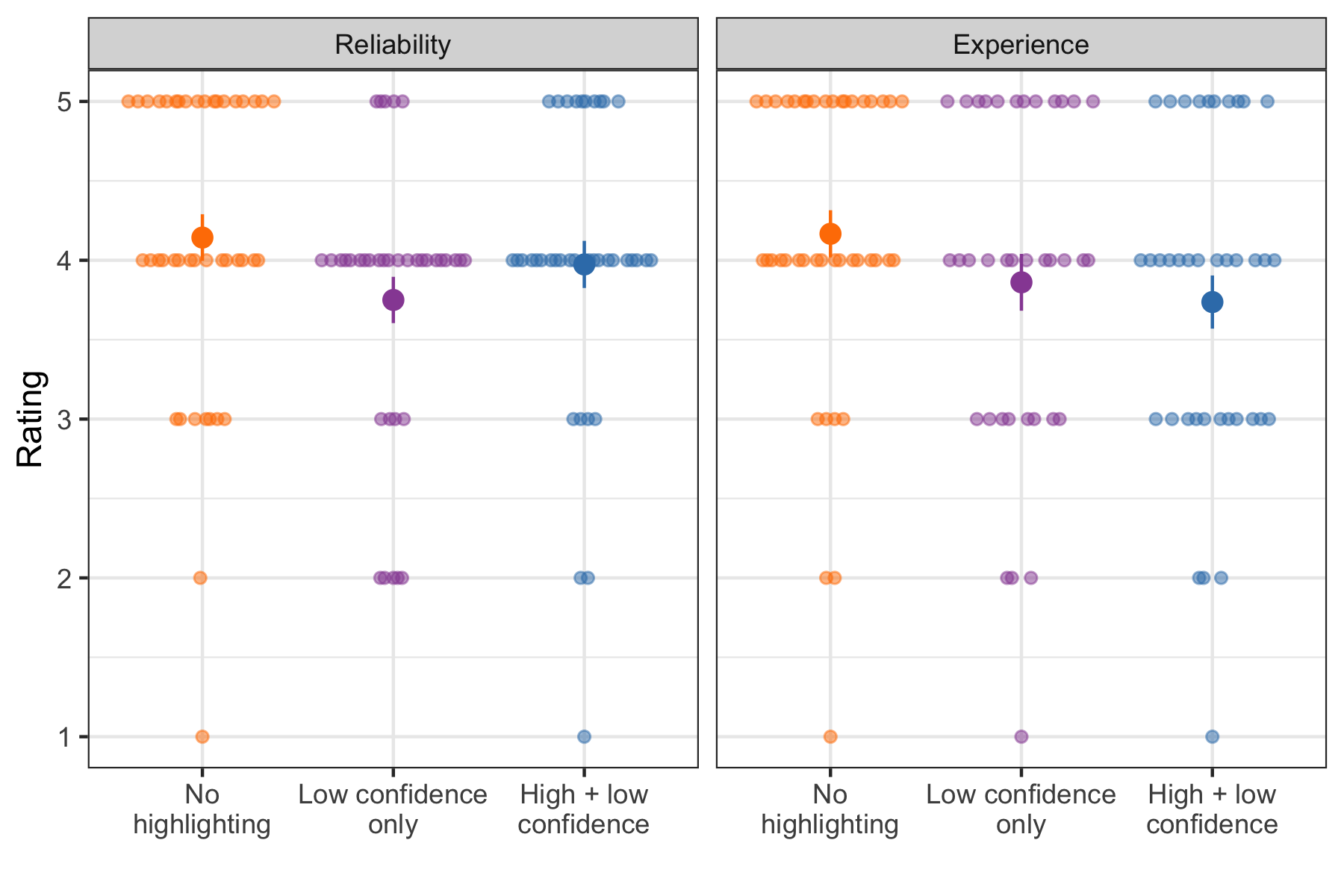}
  \caption{ Results on user perceptions (Experiment 2). Each smaller
point represents one participant's response, the larger points show the
mean by condition with error bars of +/- 1 SE. \label{fig:res_user_st_2} }
\end{figure}

\paragraph{User experience and perceived reliability.} Finally, by way of perceived reliability and search experience, we find
that all three conditions were rated quite favorably, and we detected no
systematic difference between them, as shown in Figure \ref{fig:res_user_st_2}. The no
highlighting condition had an average reliability rating of 4.1, whereas
the low confidence only condition had an average rating of 3.8 and the
high + low confidence condition had an average rating of 4.0. Neither
difference between the treatment conditions and the control is
statistically significant (low confidence vs. no highlighting: t(75.62)
= 1.90, p = 0.06; high + low confidence vs. no highlighting: t(77.70) =
0.81, p = 0.42). Similarly, the average search experience rating was 4.2
for the no highlighting condition, 3.9 for the low confidence only
condition, and 3.7 for the high + low confidence condition. Despite a
directional trend in the estimated means, neither difference is
statistically significant (low confidence vs. no highlighting: t(71.00)
= 1.31, p = 0.19; high + low confidence vs. no highlighting: t(75.75) =
1.92, p = 0.06).

\section{Discussion and Conclusion}\label{discussion-and-conclusion}

In this work, we investigated how LLM-based enhancements in search tools affect efficiency (time and number of queries), accuracy, user experience, and the ability to detect errors in a consumer search task. To obtain these measures, we created a novel experimental platform that, holding all else constant, allows participants to be randomly assigned to use traditional or LLM-based search and keeps detailed records of their interactions.

With respect to our first research question on efficiency, we found in Experiment 1 that having access to an LLM-based search tool led to a substantial increase in search efficiency. Participants using the LLM-based search tool were able to complete tasks in almost half the time compared to those using a traditional search engine. In addition, we observed a slight reduction in the number of queries issued, accompanied by a significant increase in query complexity. In other words, LLM-based search allowed people to reach decisions faster and in fewer steps by issuing queries and receiving responses that more directly addressed the decisions at hand. Regarding our third research question on perceptions, observed increases in efficiency were accompanied by significant increases in favorable ratings for the LLM-based search tool, based on participants' self-reports of their overall experience.

Our second research question concerned accuracy, a dimension on which LLMs have been known to falter. In Experiment 1, we found comparable accuracy levels between participants using LLM-based and traditional search tools for tasks that can be considered routine for the LLM, however, we found a significant drop in accuracy in a task that was difficult for the LLM, with almost half of the participants in this condition making an incorrect decision. In contrast, the vast majority of participants using traditional search tools appeared to have been able to obtain the information needed to make a correct decision. In investigating this drop in accuracy, we found that, without appropriate cues, participants using LLM-based search were overreliant on the tool, with the majority of people (60\%) making just a single query before reaching a decision. 
Even though many participants using the LLM-based search tool received incorrect information for one of the tasks, there was no significant difference in participants' subjective ratings of the reliability of the results they were shown. 

In Experiment 2 we proposed and tested mitigations for this issue of overreliance. In particular we tested whether people would compensate for erroneous LLM responses when given cues about the tool's confidence in its own responses. When LLM responses were color-coded to reflect the confidence of the model, people's accuracy increased significantly. We observed that participants were much more likely to seek further information for answers that were highlighted to reflect low confidence. Our results indicate that although automatically generating calibrated signals of model confidence is technically challenging, conveying such signals to users can effectively reduce overreliance.

Two open questions are how to best
identify potential errors in LLM-based outputs and how to best convey
confidence to users so that they can make informed decisions. In our experiments,
we used token probabilities to identify potentially incorrect responses.
In order to do that, we had to move from GPT-3.5 to GPT-3.0, because token
probabilities were not available for GPT-3.5 and later models. Token probabilities are a start but have drawbacks, for instance they are not perfectly correlated with error and are not necessarily calibrated with correctness (\cite{OpenAI2023}).  
In ongoing research, we are investing better ways to identify incorrect information and to convey the various types of errors that LLMs might make, and how to best provide cues about potential errors to users.

Another area for future research is identifying other ways that LLM-based tools can communicate 
information about the correctness of their responses, without departing from purely text-based answers. 
We incidentally found one such method when 
designing these experiments. 
Initially, we had used a meta-prompt that did not ask
the LLM to show its work---our assumption was that
presenting too many numbers may confuse a participant, increasing the
likelihood of them transcribing the wrong answer. However, we quickly realized 
that prompting the LLM to ``show its work'' provided a useful way for participants to pick up on internal inconsistencies in the
LLM-based output (an indicator of a questionable answers) and for them
to learn the basic ranges for various values (helping them better spot
anomalies). Research on numerical
perspectives has found that people are more likely to spot errors in
numbers when those numbers are put into perspective with familiar
objects (\cite{barrio_improving_2016}). One error the LLM-based tool
made in these experiments was reporting the seats up cargo space
instead of the total cargo space in an SUV, which is measured with the second and third
rows of seats down. Because the typical total cargo space in an SUV---which is usually around 16 feet long---is much larger than 16 cubic feet, comparing the latter to a familiar object could help people realize that such a number is unrealistically small.

While this investigation found many benefits of LLM-based search, it also
uncovered a strength of traditional search. In practice, it is rare for people 
to query more than one product and one
dimension at a time, despite how common it is for consumers to compare products on 
multiple dimensions (\cite{payne1993adaptive}). 
In the first experiment, participants in all conditions
were effectively encouraged to try more complex searches. Surprisingly, participants in the
traditional search condition found correct answers in fewer
than four queries on average; sometimes they found them in just one query. In our experiments this paid off because there are many pre-generated
webpages comparing particular pairs of cars on all of their dimensions, which is quite helpful for this task.
Therefore, it is possible
that the introduction of LLM-based search norms may encourage people to issue more complex
queries to traditional search engines, improving their efficiency with existing tools.

As to limitations, we explored a specific search domain 
and a simplified decision task. We focused on searches involving the purchase of high 
cost durable goods and we expect that different scenarios involving search will be affected in 
different ways. For example, while users may quickly move to more complex queries when 
doing research on purchasing a car, they may
continue to use simpler searches in other product categories. Furthermore, we expect 
users to respond differently to cues about the correctness of information under different 
conditions. For example, we observed that providing signals of the LLM's low confidence in 
certain measurements led to users doing more searches to uncover information in a car buying 
scenario. 
However, users may be inclined to accept such pieces of information in other scenarios
if they are more interested in a range than in exact numbers (for example when searching for the number of calories in a specific type of food).
We also designed our LLM-based search tool around non-conversational version of GPT for a tight experimental contrast with traditional search.
A natural avenue for future investigation would be to include conversational capabilities, and to explore tools such as Bing Chat that blend traditional search and LLMs.
Finally, we only considered a search scenario involving a choice among two vehicles, 
yet such scenarios generally involve more options, specifications, and time. We therefore just 
captured a portion of a much longer search process.

In sum, LLM-based search stands to permanently change how people search for information
online. 
The studies presented here suggest that search efficiency and satisfaction will likely increase, 
but overreliance may become more of a concern.
We hope that there
will be continued innovation and testing of ways of communicating uncertainty in AI
responses so that they may be viewed with an appropriate level of confidence. 
Uncertainty in answers, and in the world for that matter, can never be eliminated, 
but effective means of communicating it can augment human cognition and decision making.

\section*{Acknowledgements}\label{acknowledgements}

We thank Jiawei Liu for the initial development of the web framework
used in this experiment. We thank Mark Whiting and Duncan Watts for
assistance in recruiting participants in the second experiment. We thank
the audience at University of Pennsylvania's ``Large Language Models:
Behavioral Science Meets Computer Science'' workshop on May 19, 2023, and Microsoft Research's ``AI, Cognition, and the Economy workshop'' on October 12, 2023. We also thank Susan Dumais for her invaluable feedback.

\printbibliography

\appendix

\renewcommand\thefigure{\thesection.\arabic{figure}}    
\setcounter{figure}{0}   

\setcounter{table}{0}
\renewcommand{\thetable}{A\arabic{table}}

\newpage
\section{Appendix}

\subsection{Traditional Search Activity}

The following table  provides a breakdown of Bing searches for top SUVs in terms of the number of vehicles and dimensions in the first half of 2022.

\begin{table}[ht]
\resizebox{\textwidth}{!}{%
\begin{tabular}{@{}cccc@{}}
\toprule
\textbf{Number of Vehicles} & \textbf{Number of Dimensions} & \textbf{Percent of Queries} & \textbf{Percent of Queries (w/ 1+ dimension)} \\
\midrule
1                  & 0                    & 63.00\%            & –                                    \\
1                  & 1                    & 25.18\%            & 68.40\%                              \\
1                  & 2                    & 9.22\%             & 25.05\%                              \\
1                  & 3                    & 2.00\%             & 5.43\%                               \\
1                  & 4                    & 0.23\%             & 0.63\%                               \\
1                  & 5                    & 0.01\%             & 0.01\%                               \\
1                  & 6                    & 0.00\%             & 0.00\%                               \\
2                  & 0                    & 0.18\%             & –                                    \\
2                  & 1                    & 0.10\%             & 0.26\%                               \\
2                  & 2                    & 0.07\%             & 0.18\%                               \\
2                  & 3                    & 0.01\%             & 0.04\%                               \\
2                  & 4                    & 0.00\%             & 0.00\%                               \\
3                  & 0                    & 0.00\%             & –                                    \\
3                  & 1                    & 0.00\%             & 0.00\%    \\              \bottomrule  
\end{tabular}%
}
\caption{Table of number of products and dimensions in all searches for top 25 SUV in 2022. Starting with a list of the top 25 SUV by sales in the first half of 2022  we looked at every 2022 Bing search that included these 25 SUVs and and the top 10 most queried dimensions (e.g., cargo space, length, etc). Most queries mention only one vehicle. If a dimension is mentioned, most queries mention only one dimension.
\label{tab:products}}
\end{table}

\subsection{Experiment 1}

\subsubsection{Tutorial}

\begin{figure}[h!]
  \centering
  \includegraphics[width=0.7\linewidth]{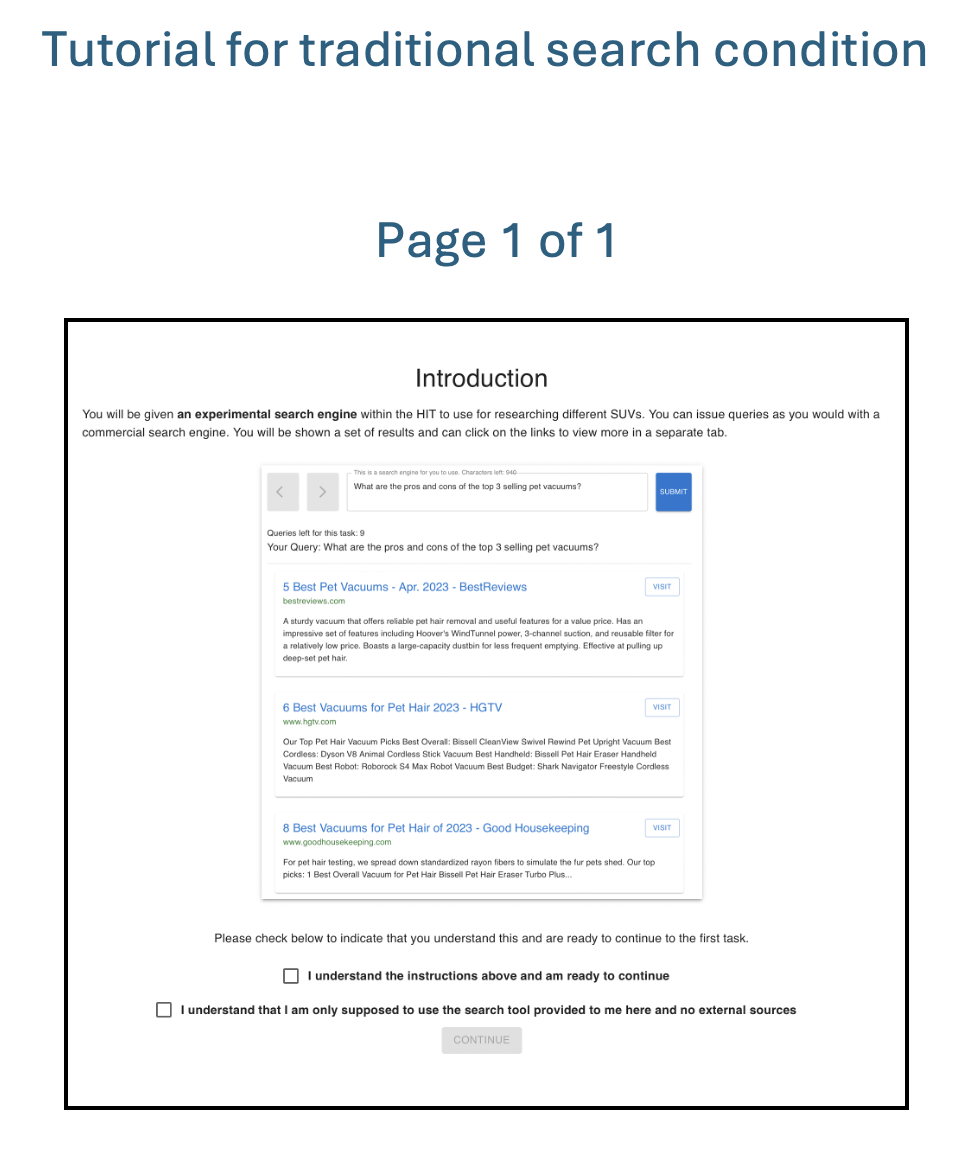}
  \caption{ The tutorial for participants who were in the
traditional search condition (Experiment 1). \label{fig:app_tutorial_search} }
\end{figure}

\begin{figure}[h!]
  \centering
  \includegraphics[width=\linewidth]{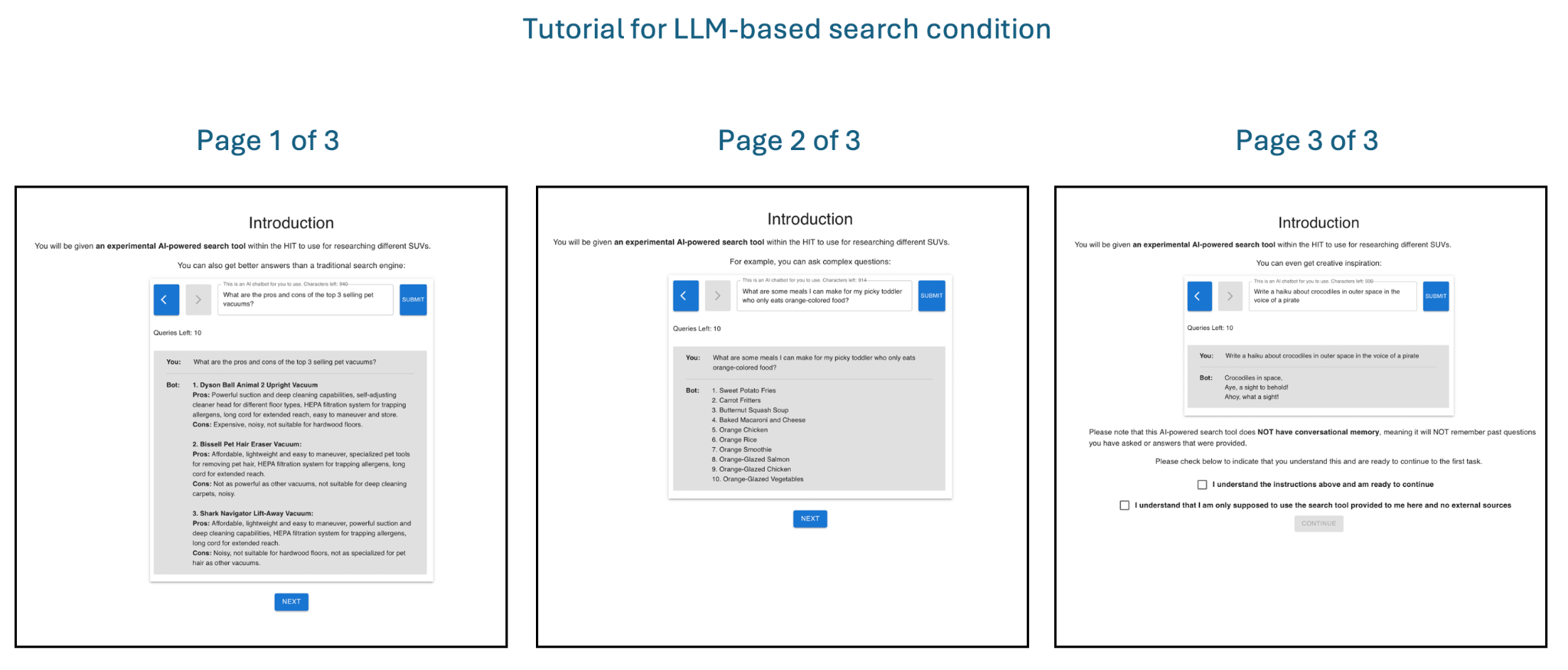}
  \caption{ The tutorial for participants who were in the
LLM-based search condition (Experiment 1).\label{fig:app_tutorial_llm} }
\end{figure}

Before starting the first task, participants
in each condition were given a short tutorial on what to expect from the
search tool they would be using. Figure \ref{fig:app_tutorial_search} shows the tutorial for the
traditional search condition and Figure \ref{fig:app_tutorial_llm} shows the tutorial for
LLM-based search.

\subsubsection{Speed and accuracy jointly}

\begin{figure}[h!]
  \centering
  \includegraphics[width=\linewidth]{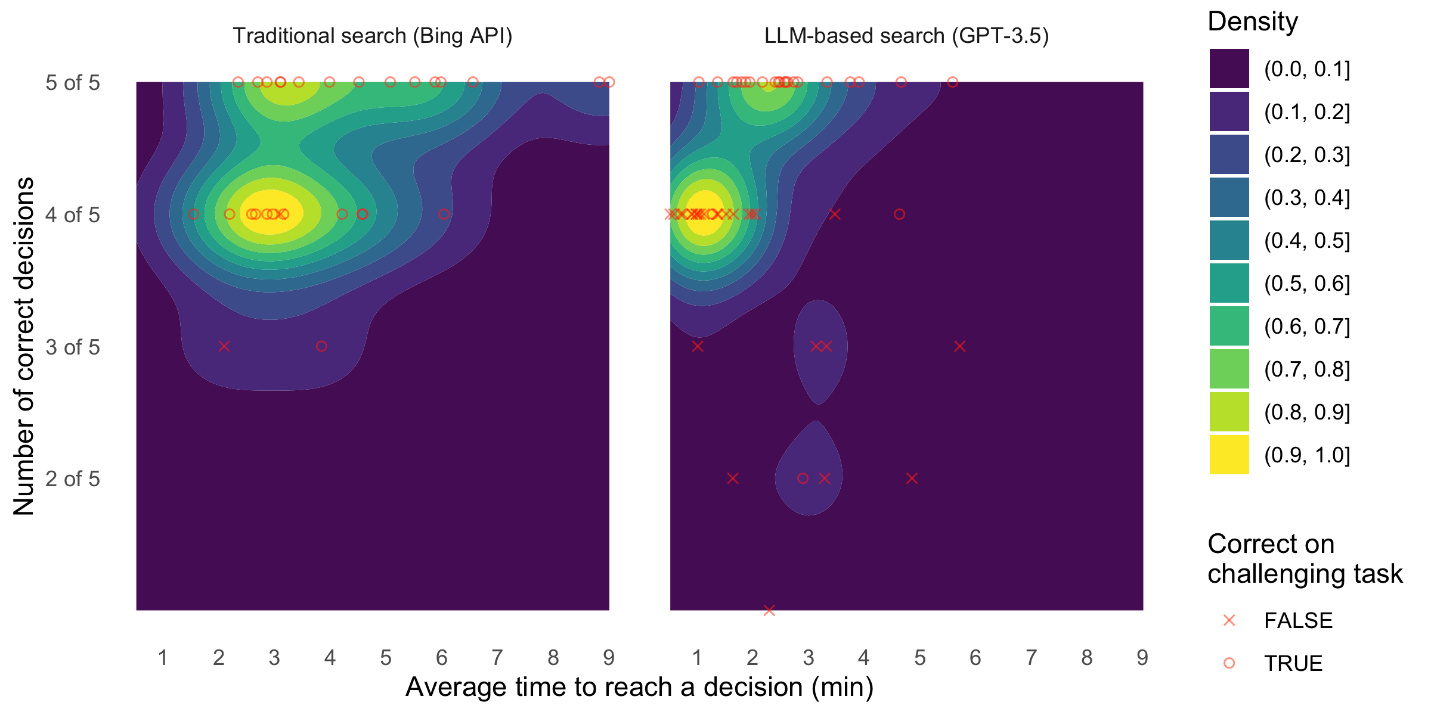}
  \caption{Joint view of speed and accuracy (Experiment 1). Each point
represents the data from one participant over five questions. Points are
represented with an ``o'' if they got the challenging question correct
and a ``x'' if they failed to.\label{fig:app_speed_acc} }
\end{figure}

\paragraph{Speed and accuracy jointly.} Speed and accuracy are both
desirable for search engine users. Figure \ref{fig:app_speed_acc} plots them against each
other. The upper left corner of each panel represents the best
performance, that is, the most correct answers in the least amount of
time. To facilitate seeing patterns despite overplotting, a density was
fit to the responses to create a heat map. The high density areas in
both panels show the participants with the LLM tools in a favorable
position near the upper left. They have less variance in time taken but
more variance in accuracy, mostly owing to the additional item designed
to be difficult. Performance on this item is marked with an x showing
that the vast majority of participants who did not score all the
questions correctly made an error on this item.

\subsection{ Experiment 2}

\subsubsection{Efficiency}

\begin{figure}[h!]
  \centering
  \includegraphics[width=\linewidth]{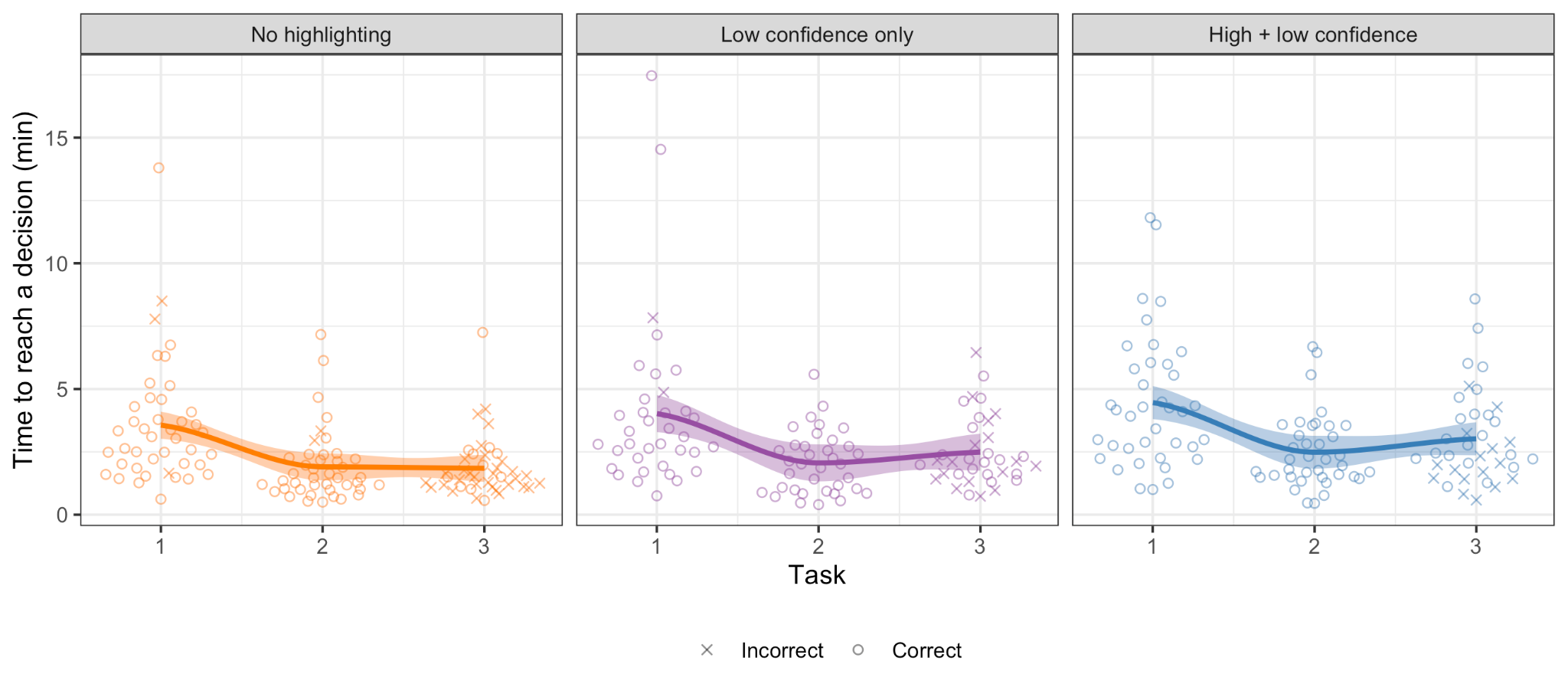}
  \caption{ Time to reach a decision in Experiment 2 by
condition and task. Each point represents one participant's number of
queries for the task. \label{fig:app_time_dec_st2} }
\end{figure}

As in our first experiment, across
conditions we see a learning effect where participants take less time to
reach a decision on the second task compared to the first (Figure \ref{fig:app_time_dec_st2}).
Using a similar linear mixed model as in Experiment 1 to model log task
duration on the routine tasks, we find that on average across all
conditions, participants take 3.3 minutes (95\% CI {[}2.9 minutes, 3.7
minutes{]}) to complete the first task, but only 1.8 minutes (95\% CI
{[}1.6 minutes, 2.0 minutes{]}) to complete the second task. Averaged
over both of these tasks, we find that participants in the treatment
conditions were slightly slower than those in the control overall, with
a statistically significant difference for high + low confidence
highlighting compared to no highlighting (t(113) = 2.09, p = 0.04) but
no evidence of a systematic difference for low confidence highlighting
only (t(113) = 0.63, p = 0.53). On the third task, where participants
encounter potentially unreliable information, we see an increase in time
to decision for the two treatment conditions that highlight potentially
unreliable information, but no such increase for the control without
confidence highlighting (low confidence only vs. no highlighting:
t(72.79) = -2.53, p = 0.01; high + low confidence vs. no highlighting:
t(72.63) = -3.70, p \textless{} 0.001).

\begin{figure}[t!]
  \centering
  \includegraphics[width=\linewidth]{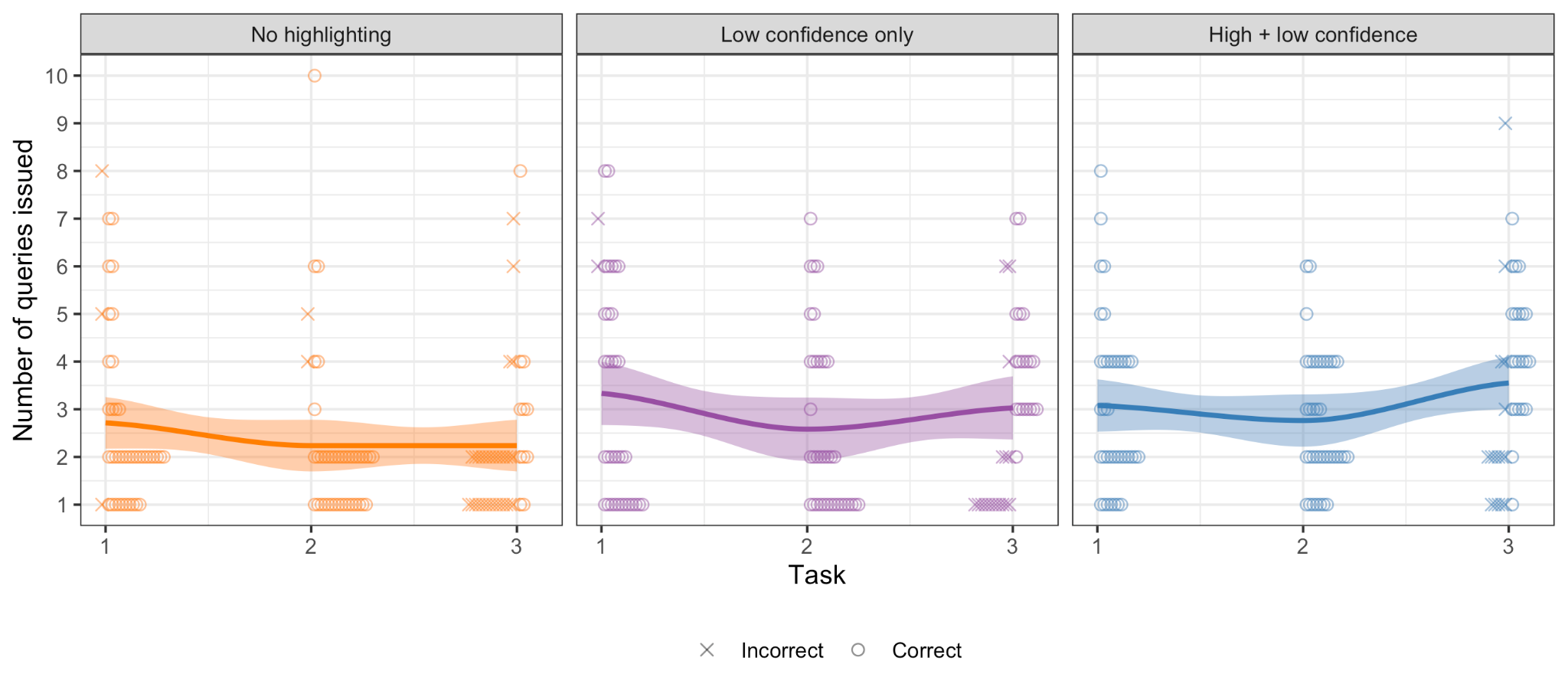}
  \caption{ Number of queries issued in Experiment 2 by
condition and task. Each point represents one participant's number of
queries for the task.\label{fig:app_queries_st2} }
\end{figure}

Analyzing the number of queries using a similar linear mixed model as in
Experiment 1, we find no evidence of systematic differences in the
number of queries issued in the first two routine tasks across
conditions, with participants issuing 2.3, 2.7, and 2.7 queries per task
on average for the no highlighting, low confidence only, and high + low
confidence conditions, respectively. However, in the third task we see
substantial increases in the number of queries for the two treatment
conditions compared to the control (low confidence only vs. no
highlighting: 3.0 vs 2.2 queries on average, t(70.97) = -2.00, p = 0.05;
high + low confidence vs. no highlighting: 3.6 vs 2.2 queries on
average, t(73.21) = -3.29, p = 0.002). This is visually apparent in
Figure \ref{fig:app_queries_st2}, as depicted in uptick in queries for the middle and right
panels compared to the left panel.

\end{document}